\documentclass[12pt]{article}
\usepackage{epsfig}
\newcommand{\eps}{\varepsilon}
\newcommand{\slq}{q\kern-5.5pt/}
\newcommand{\slv}{v\kern-5pt\raise1pt\hbox{$\scriptstyle/$}\kern1pt}
\newcommand{\pfrac}[2]{\left(\frac{#1}{#2}\right)}
\newcommand{\Li}{{\rm Li}}
\newcommand{\dDp}{\frac{d^Dp}{(2\pi)^D}}
\newcommand{\MeV}{{\rm\,MeV}}
\newcommand{\GeV}{{\rm\,GeV}}

\begin{document} 

\begin{flushright}
MZ-TH/00-24\\
hep-ph/0006072\\
June 2000
\end{flushright}

\vspace{1truecm}

\begin{center}
{\Large\bf QCD sum rules as applied to heavy baryons\footnote{Invited talk
  given at the conference ``Heavy Quark Physics 5'', Dubna, Russia, 6--8 April
  2000, to appear in the proceedings}}\\[1truecm]
{\Large Stefan Groote}\\[.7truecm]
{\large Institut f\"ur Physik der\\[.2truecm]
  Johannes-Gutenberg-Universit\"at,\\[.3truecm]
  55099 Mainz, Germany}
\end{center}

\vspace{1truecm}

\begin{abstract}
We give an overview over recent calculations of baryonic correlator functions
with finite mass quarks in view on their applicability for QCD sum rules. The
QCD sum rule method is then demonstrated within the Heavy Quark Effective
Theory.
\end{abstract}

\newpage

\section{Mesonic and baryonic correlators}
Data sets which are produced to an huge amount especially by the so-called
$B$ factories like Barbar, Belle, Cleo and Hera B allow for the discovery of
excited states of hadrons containing the $b$ quark and other heavy quarks. In
order to follow this development on the experimental side, theorists are asked
to develop and use methods to analyze such excited states. There is one main
obstacle on this way which consists of the scale differences that occur. While
perturbative calculations including a heavy quark can effectively be done only
in the high energy range, there is an extrapolation of these results down to
the energy range needed where we expect excited states are to be located.

One of the most powerful methods which was developed long ago are the QCD
sum rules. In this talk we will only concentrate on the QCD sum rule approach
developed by Shifman, Vainshtein and Zakharov, called SVZ
approach~\cite{ShifmanVainshteinZakharov}. This approach makes special
assumptions about the spectral density of the correlator function related to
the hadron and uses the Borel transform for the extrapolation. There are a lot
of calculations for the correlator function of hadrons, especially for those
containing a heavy quark. The calculations were performed within perturbative
QCD as well as HQET and the massless limit. In the first part of this talk I
will concentrate on calculations using perturbative QCD. In all cases the
baryons are a kind of ``stepchild'' of the theorists, so we work hard to fill
this gap.

\subsection{Baryonic correlators in QCD}
The mesonic correlator function for two vector currents has already been
calculated ten years ago by Generalis~\cite{Generalis} even in the case of two
different and non-vanishing masses. Our aim is therefore to extend these
calculations to the baryonic case. A first step has already been done by
a recent publication~\cite{GrooteKoernerPivovarov}. In this publication the
spectral density is calculated for a three-quark current (two massless and one
of finite mass) of the form
\begin{equation}
J_B=[u^{iT}Cd^j]\Psi^k\eps_{ijk}.
\end{equation}
The result presented in~\cite{GrooteKoernerPivovarov} is the one for the mass
part $\Pi_m(-q^2)$ of the correlator
\begin{equation}
\Pi(-q^2)=\slq\Pi_q(-q^2)+m\Pi_m(-q^2).
\end{equation}
This part is independently interesting because it can directly be compared
with the result obtained within HQET, as we will see later. The momentum part
$\Pi_q(-q^2)$ is still under construction, we expect a publication in the next
few months. As mentioned before, we can easily reconstruct the correlator
function
\begin{equation}
\Pi_m(-q^2)=\int_{m^2}^\infty\frac{\rho_m(s)ds}{s-q^2}
\end{equation}
if we know the spectral density. This is given by
\begin{equation}
\rho_m(s)=\frac1{128\pi^4}\rho(s),\qquad
\rho=s^2\left\{\rho_0\left(1+\frac{\alpha_s}\pi\ln\pfrac{\mu^2}{m^2}
  \right)+\frac{\alpha_s}\pi\rho_1\right\}
\end{equation}
where
\begin{equation}
\rho_0(s)=1+9z-9z^3-z^3+6z(1+z)\ln z
\end{equation}
and
\begin{eqnarray}\label{corr1}
\lefteqn{\rho_1(s)\ =\ 9+\frac{665}9z-\frac{665}9z^2-9z^3
  -\left(\frac{58}9+42z-42z^2-\frac{58}9z^3\right)\ln(1-z)\,+}\nonumber\\&&
  +\left(2+\frac{154}3z-\frac{22}3z^2-\frac{58}9z^3\right)\ln z
  +4\left(\frac13+3z-3z^2-\frac13z^3\right)\ln(1-z)\ln z\,+\nonumber\\&&
  +12z\left(2+3z+\frac19z^2\right)\left(\frac12\ln^2z-\zeta(2)\right)
  +4\left(\frac23+12z+3z^2-\frac13z^3\right)\Li_2(z)\,+\nonumber\\&&
  +24z(1+z)\left(\Li_3(z)-\zeta(3)-\frac13\Li_2(z)\ln z\right).
\end{eqnarray}
This result is obtained by using basic integrals of the kind
\begin{equation}
V(\alpha,\beta;q^2/m^2)=\int\dDp\frac1{(p^2+m^2)^\alpha(q-p)^{2\beta}}
\end{equation}
which are a generalization of the standard object $G(\alpha,\beta)$ of the
massless calculation.

\begin{center}\begin{figure}[ht]\begin{center}
\epsfig{figure=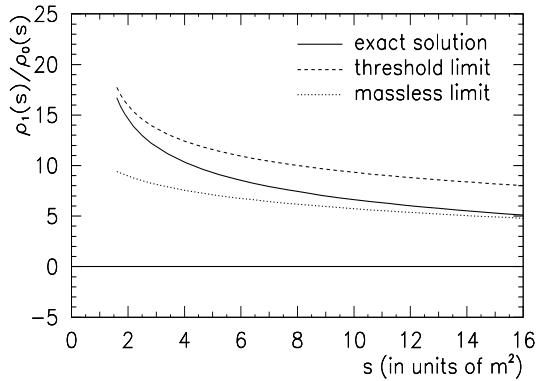, height=5truecm, width=7truecm}
\caption{\label{fig1}The ratio $\rho_1(s)/\rho_0(s)$ of the next-to-leading
correction and the leading order term in dependence of the energy square $s$}
\end{center}\end{figure}\end{center}

\subsection{Comparison with the limits}
Here we only want to stress that we were able to compare this result
with the two limits, i.e.\ the massless limit and the heavy quark or
near-threshold limit. For the massless limit we obtain
\begin{equation}
m\rho(s)=m_{\overline{\rm MS}}(\mu)s^2\left\{1+\frac{\alpha_s}\pi
  \left(2\ln\pfrac{\mu^2}s+\frac{31}3\right)\right\}
\end{equation}
where the relation between the pole (or invariant) mass parameter $m$ and the
$\overline{\rm MS}$ mass $m_{\overline{\rm MS}}$ reads
\begin{equation}
m=m_{\overline{\rm MS}}\left\{1+\frac{\alpha_s}\pi
  \left(\ln\pfrac{\mu^2}{m^2}+\frac43\right)\right\}.
\end{equation}
With this explicit finite mass representation we can even obtain terms like
$m^2\ln(\mu^2/m^2)$ which are absent in the effective theory of massless
quarks but result from perturbative contributions of heavy quark condensates
$\langle\bar\Psi\Psi\rangle$. For the other limit, i.e.\ the near-threshold
limit $E\rightarrow 0$ with $s=(m+E)^2$ we obtain
\begin{equation}\label{hqet}
\rho^{\rm thr}(m,E)=\frac{16E^5}{5m}\left\{1+\frac{\alpha_s}\pi
  \ln\pfrac{\mu^2}{m^2}+\frac{\alpha_s}\pi\left(\frac{54}5+\frac{4\pi^2}9
  +4\ln\pfrac{m}{2E}\right)\right\}+O(E^6).
\end{equation}
The invariant function $\rho_m$ suffices to determine the complete leading
HQET behaviour since one has $\slq\rho_q+m\rho_m\rightarrow m(\slv+1)\rho$ for
the leading term. In this region the appropriate device to compute the limit
of the correlator is HQET (see e.g.~\cite{Georgi,Neubert}). Writing
\begin{equation}
m\rho^{\rm thr}(m,E)=C(m/\mu,\alpha_s)^2\rho^{\rm HQET}(E,\mu)
\end{equation}
we obtain the known result for
$\rho^{\rm HQET}(E,\mu)$~\cite{GrooteKoernerYakovlev1} with matching
coefficient $C(m/\mu,\alpha_s)$ \cite{GrozinYakovlev}. In this case the
matching procedure allows one to restore the near-threshold limit of the full
correlator starting from the simpler effective theory near
threshold~\cite{EichtenHill} (see also~\cite{thresh}). Note that the higher
order corrections in $E/m$ to Eq.~(\ref{hqet}) can be easily obtained from the
explicit result given in Eq.~(\ref{corr1}). Indeed, the next-to-leading order
correction in low energy expansion reads
\begin{equation}
\Delta\rho^{\rm thr}(m,E)=-\frac{88E^6}{5m^2}
  \left\{1+\frac{\alpha_s}\pi\left(\ln\pfrac{\mu^2}{m^2}
  +\frac{376}{33}+\frac{4\pi^2}9+\frac{140}{33}\ln\pfrac{m}{2E}\right)\right\}.
\end{equation}
It is a much more difficult task to obtain this result starting from HQET. 
In Fig.~\ref{fig1} we compare components of the baryonic spectral function in
leading and next-to-leading order. Shown is the ratio $\rho_1(s)/\rho_0(s)$
where we put $\mu=m$ for simplicity. One can see that a simple interpolation
between the two limits can give a rather good approximation for the
next-to-leading order correction in the complete region of $s$.

\begin{center}\begin{figure}[ht]\begin{center}
\epsfig{figure=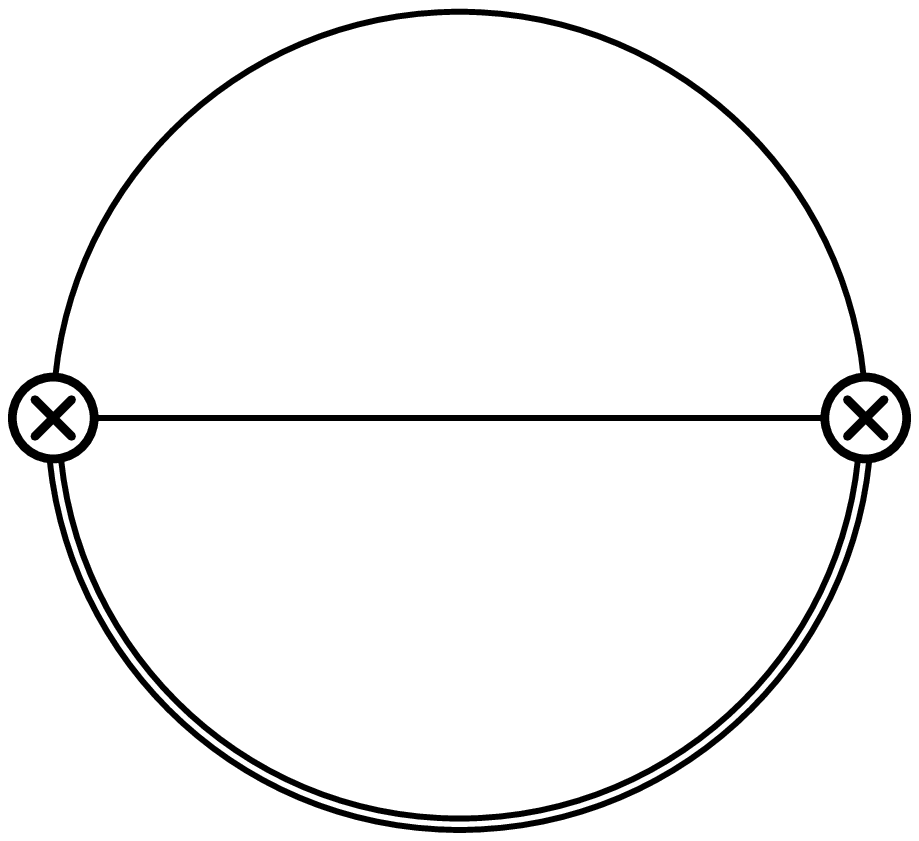, height=2truecm, width=2truecm}\kern12pt
\epsfig{figure=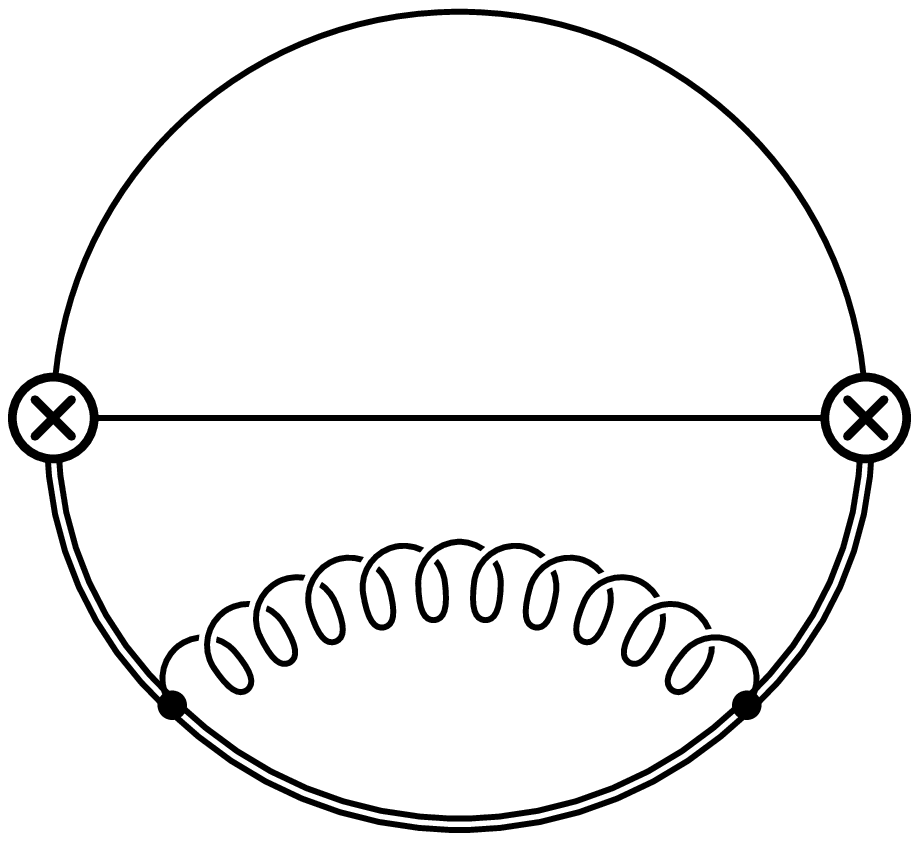, height=2truecm, width=2truecm}\kern12pt
\epsfig{figure=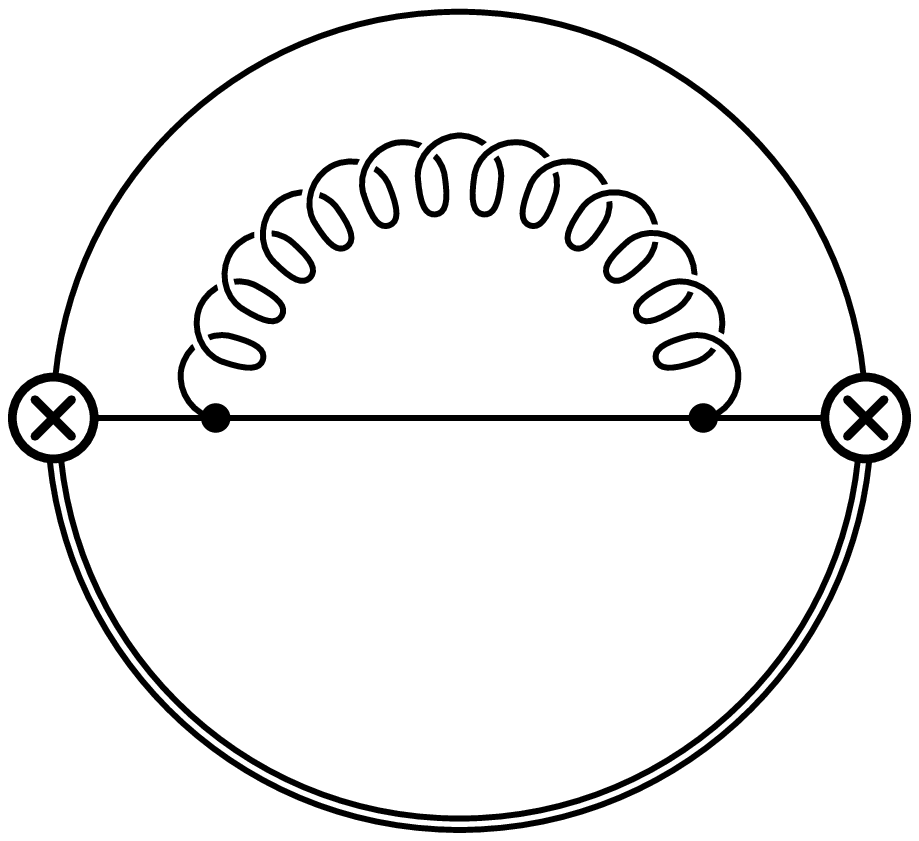, height=2truecm, width=2truecm}\kern12pt
\epsfig{figure=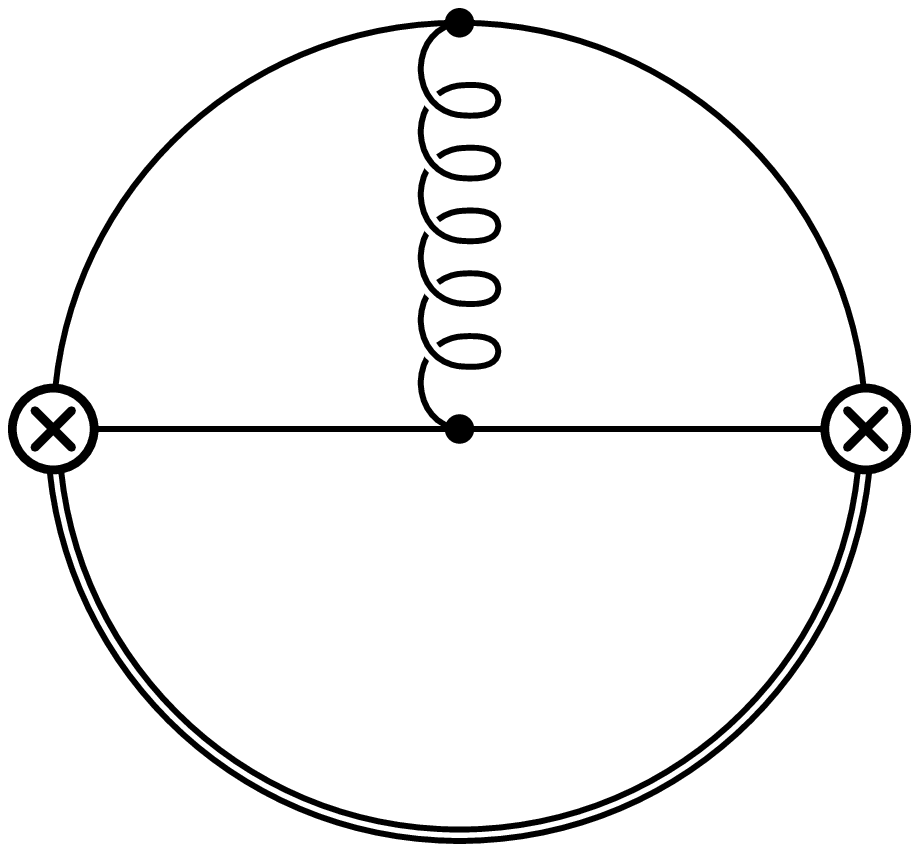, height=2truecm, width=2truecm}\kern12pt
\epsfig{figure=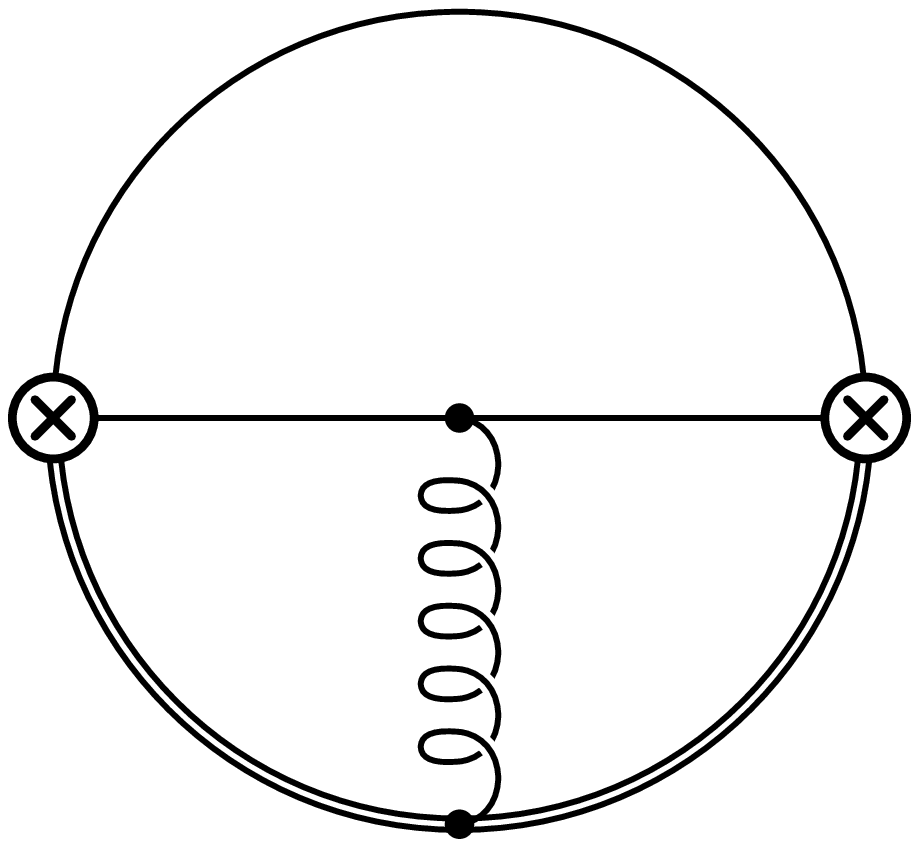, height=2truecm, width=2truecm}\end{center}
\caption{\label{fig2}The calculated two-loop and three-loop topologies}
\end{figure}\end{center}

\subsection{A specific feature of the calculation}
\begin{center}\begin{figure}[ht]\begin{center}
\epsfig{figure=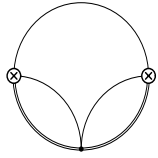, height=2truecm, width=2truecm}\end{center}
\caption{\label{fig2a}The scalar spectacle diagram}
\end{figure}\end{center}

The two-loop and three-loop Feynman diagrams which had to be calculated
for the determination of the spectral function are shown in Fig.~\ref{fig2}.
For our calculation we extensively used the fact that the diagrams can be
composed by glueing together subdiagrams (see also Ref.~\cite{melone}).
The so-called scalar spectacle diagram in Fig.~\ref{fig2a} is calculated as
the convolution of two heavy-light spectral densities,
\begin{equation}
\rho(s)=\int\lambda(s,s_1,s_2)\rho_1(s_1)\rho_2(s_2)ds_1ds_2
\end{equation}
where the convolution function is given by the remaining line.

\section{QCD sum rules for the HQET}
The QCD sum rules for the correlator calculated above have not yet been
constructed. Instead we will present the principles of the SVZ approach to QCD
sum rules as applied to leading order HQET with results for heavy baryons
taken from previous
publications~\cite{GrooteKoernerYakovlev1,GrooteKoernerYakovlev2}.

\subsection{Construction of QCD sum rules}
The QCD sum rules can be constructed by taking care of two possible
expressions for the two-point correlator which in this case reduces to a
scalar correlator function $P(\omega)$,
\begin{equation}
\Pi(\omega=p\cdot v)=i\int e^{ip\cdot x}
  \langle T\{J(x),\bar J(0)\}\rangle d^4x
  =\Gamma'\frac{1+\slv}2\bar\Gamma'\frac12{\rm Tr}(\Gamma\bar\Gamma)
  2{\rm\,Tr}(\tau\tau^\dagger)P(\omega).
\end{equation}
On the one hand side, the function $P(\omega)$ satisfies the dispersion
relation
\begin{equation}
P(\omega)=\int_0^\infty\frac{\rho(\omega')d\omega'}{\omega'-\omega-i0}
  +\mbox{subtraction},
\end{equation}
where $\rho(\omega)$ is the spectral density in HQET. On the phenomenological
side the two-point correlator is represented by the spectral representation
\begin{equation}
P(\omega)=\frac{\frac12|F_B|^2}{\bar\Lambda-\omega-i0}
  +\sum_{X\ne B}\frac{\frac12|F_X|^2}{\omega_X-\omega-i0}
  +\mbox{subtr.}
\end{equation}
where $\bar\Lambda=m_B-m_Q$ is the ground state energy of the baryon and
$F_B$ the residue. The main assumption of the SVZ approach is that the
remaining sum can be approximated by the integral of the spectral density
given by the dispersion relation and starting from some threshold energy
$E_C$. The combination of the phenomenological and the theoretical identity
for the correlator function then leads to
\begin{equation}
\frac{\frac12|F_B|^2}{\bar\Lambda-\omega-i0}=\int_0^{E_C}
  \frac{\rho(\omega')d\omega'}{\omega'-\omega-i0}+\mbox{subtr.}
\end{equation}

\subsection{The Borel transformation}
This formula is not useful since the spectral density as calculated in the
Euklidean domain is reliable only for negative values of $\omega$, while the
integral is to be calculated mainly at $\omega=\bar\Lambda$. This region of
integration can be reached by an extrapolation using higher and higher
derivatives when $\omega$ goes to $-\infty$. This extrapolation is expressed
by the Borel transformation (cf.\ e.g. Ref.~\cite{Neubert})
\begin{equation}
\hat B_T^{(\omega)}(f(\omega))=\lim_{-\omega,n\rightarrow\infty}
  \frac{(-\omega)^{n+1}}{n!}\frac{d^n}{d\omega^n}f(\omega),\qquad
  T=\frac{-\omega}n\quad\mbox{fixed}.
\end{equation}
The Borel transformation is by construction a derivative and therefore also
cancels the (constant) subtraction terms. The Borel parameter $T$ is an
unphysical quantity in units of an energy, and the obtained values should be
mostly independent on this parameter. This will be the main criterion in
analyzing the sum rules. The Borel transformation leads to the final form of
the QCD sum rule,
\begin{equation}\label{sumrule1}
\frac12|F_B|^2e^{-\bar\Lambda/T}=K(E_C,T),\qquad K(E_C,T)=\int_0^{E_C}
  \rho(\omega)e^{-\omega/T}d\omega+\sum_i\frac{A_i}{T^i}.
\end{equation}
At the end we can take the derivative with respect to the inverse Borel
parameter and obtain a second sum rule,
\begin{equation}\label{sumrule2}
\bar\Lambda=-\frac12\frac\partial{\partial(T^{-1})}\ln K(E_C,T).
\end{equation}

\begin{center}\begin{figure}[ht]\begin{center}
\epsfig{figure=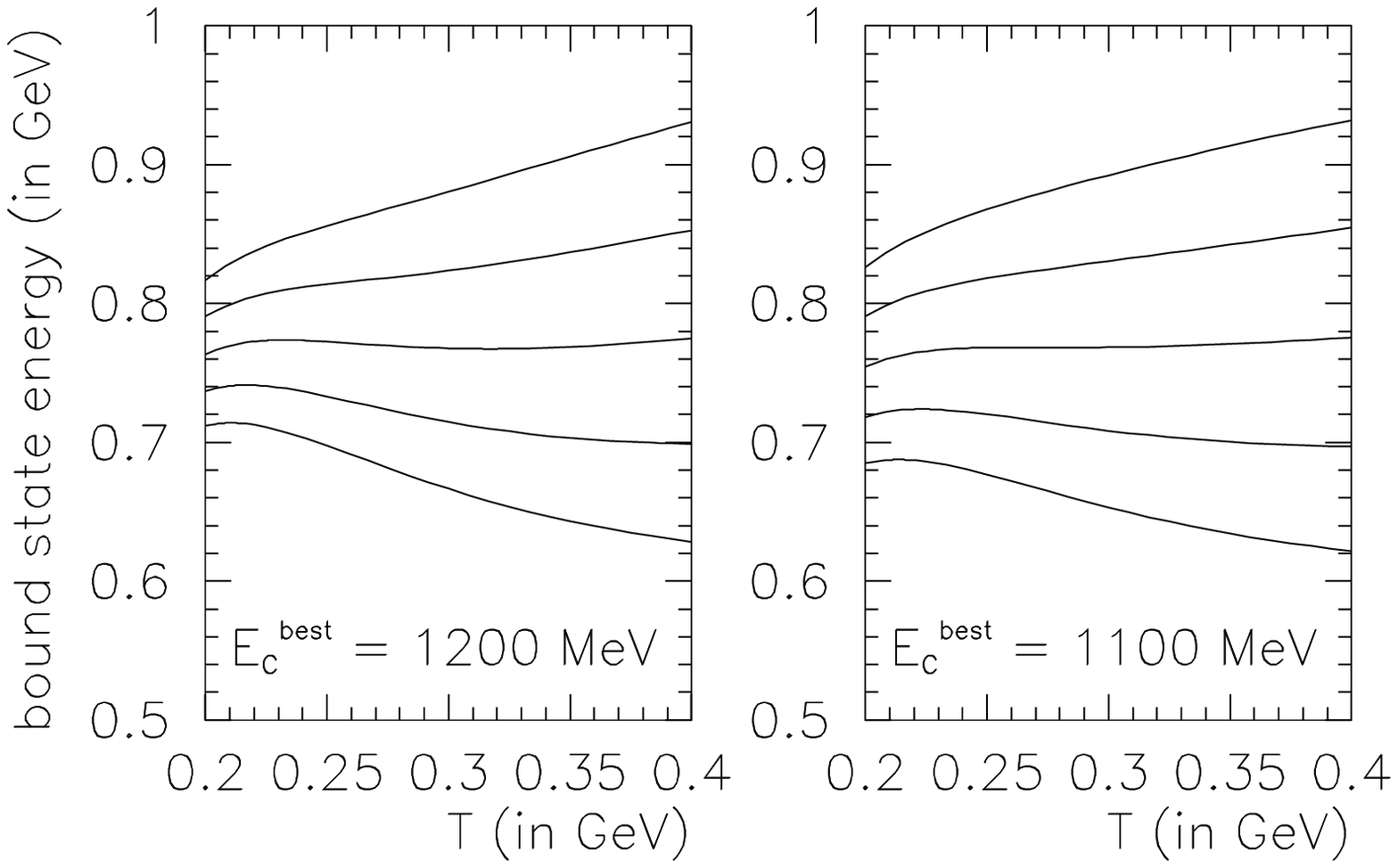, height=5truecm, width=7truecm}
\epsfig{figure=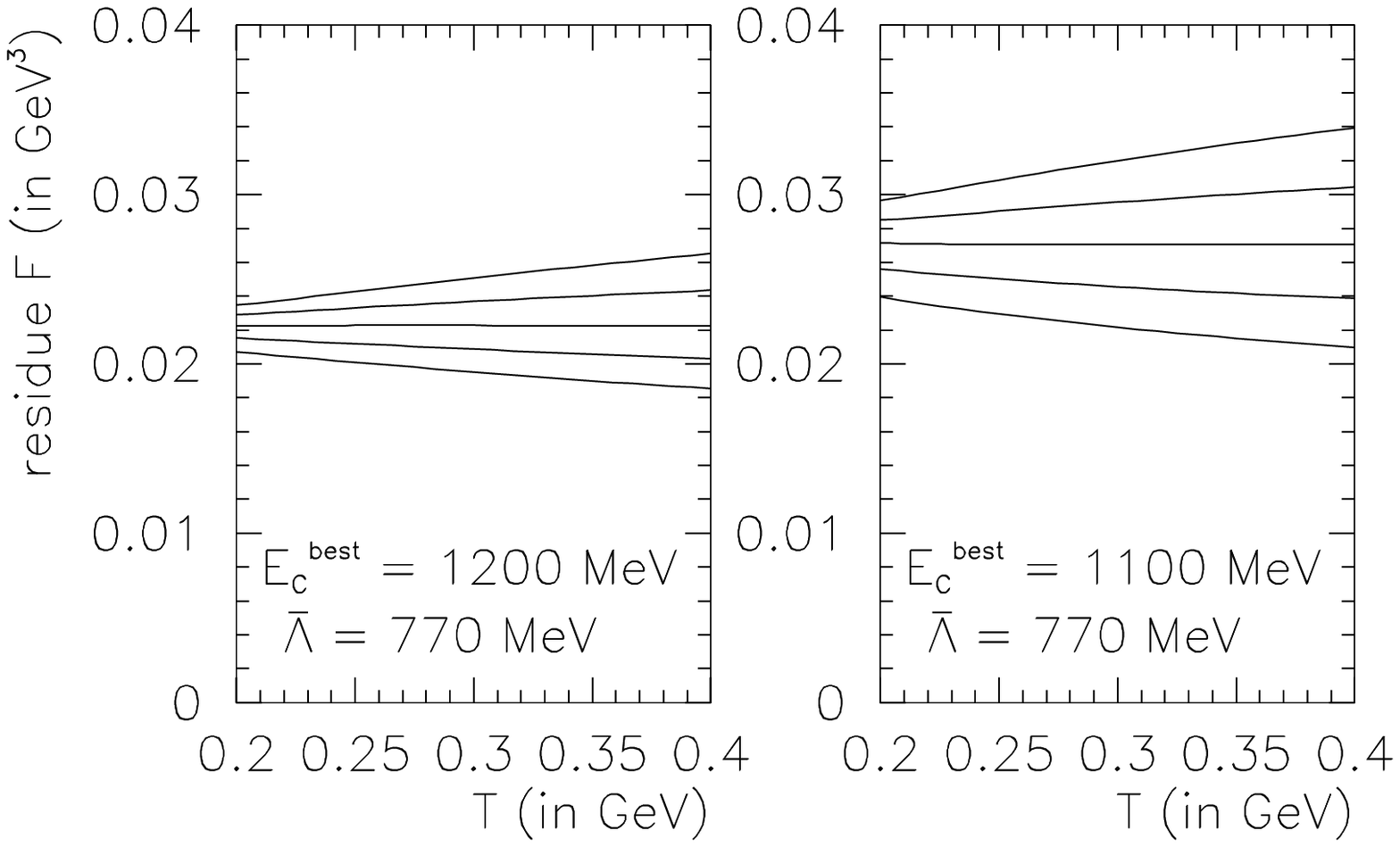, height=5truecm, width=7truecm}
\hbox{\bf(a)\kern3truecm(b)\kern3truecm(c)\kern3truecm(d)}\end{center}
\caption{\label{fig3}Sum rule analysis for the heavy baryon $\Lambda_Q$.
Shown are five curves for five different values of the threshold energy $E_C$
spaced by $100\MeV$ around the central value $E_C=E_C^{\rm best}$. $E_C$
grows from bottom to top. The figure parts are in detail (a) the lowest order
sum rule results and (b) the $O(\alpha_s)$ sum rule results for the bound state
energy $\bar\Lambda$, (c) the lowest order sum rule results and (d) the
$O(\alpha_s)$ sum rule results for the absolute value of the residue
$F_\Lambda$.}
\end{figure}\end{center}

\subsection{QCD sum rule analysis}
The procedure of the sum rule analysis is as follows: First we calculate the
theoretical expression for $K(E_C,T)$ by calculating the spectral density
within perturbation theory. This has been done for the next-to-leading order
of the $d=0$ term as well as the $d=3$ term in the Operator Product Expansion.
For brevity I will only mention the results of the analysis for the $d=0$
radiative corrections. For the analysis we have to select a ``sum rule
window'', i.e.\ a range for the Borel parameter $T$ in which the sum rule
analysis is performed. The boundaries of this window is given by heuristic
arguments. If the Borel parameter becomes too small, the nonperturbative
contributions which are badly known blow up. If on the other hand the Borel
parameter becomes too large, the fact that the Borel parameter appears as
``temperature'' in the sum rule causes the problem that higher and higher
excitations contribute to the ground state. The region of reliability and so
the sum rule window is therefore roughly given by
$\Lambda_{\rm QCD}<T<2\bar\Lambda$.

Now we use the second sum rule~(\ref{sumrule2}) to determine the ground state
energy parameter $\bar\Lambda$. This is done by varying the continuum
threshold parameter $E_C$ in order to obtain a rather stable value for the
quantities with respect to the unphysical parameter $T$ within the sum rule
window. The value obtained for the ground state energy can then be used in the
first sum rule~(\ref{sumrule1}) to determine the absolute value $|F_B|$ of the
residue. More detailed considerations are found in
Ref.~\cite{GrooteKoernerYakovlev1}. Our results read
\begin{eqnarray}
E_C(\Lambda_Q)&=&1.2\pm 0.1\GeV,\nonumber\\
\bar\Lambda(\Lambda_Q)&=&0.77\pm 0.05\GeV,\\
|F(\Lambda_Q)|&=&0.022\pm 0.001\GeV^3\nonumber
\end{eqnarray}
while the $O(\alpha_s)$ results are given by
\begin{eqnarray}
E_C(\Lambda_Q)&=&1.1\pm 0.1\GeV,\nonumber\\
\bar\Lambda(\Lambda_Q)&=&0.77\pm 0.05\GeV,\\
|F(\Lambda_Q)|&=&0.027\pm 0.001\GeV^3\nonumber
\end{eqnarray}
where the errors are estimated by looking at the stability with respect to
different values for $E_C$. Taking the experimental results for the masses of
the baryons, namely $m(\Lambda_c)=2284.9\pm 0.6\MeV$ and
$m(\Lambda_b)=5642\pm 50\MeV$~\cite{PDG}, our central value
$\bar\Lambda(\Lambda_Q)$ for the bound state energy suggests pole masses of
$m_c=1520\pm 100\MeV$ and $m_b=4880\pm 100\MeV$.

\section{Conclusion and Outlook}
The QCD sum rule analysis for heavy baryons requires the calculation of the
spectral density related to the correlator function of baryonic currents. If
this spectral density is calculated, the sum rule analysis can be used as
powerful tool to extract phenomenological non-perturbative quantities like
bound state energies and form factors. I have presented the sum rule analysis
for an example within HQET. In order to perform a sum rule analysis for
baryons containing finite mass quarks the following steps are still missing
and will be done in the near future:
\begin{itemize}
\item The calculation of the momentum part of the spectral density
  (nearly finished)
\item The matching procedure for the spectral density
\item The calculation of spectral densities for baryons with different
  quantum numbers
\end{itemize}
Moreover it is planned to develop light-cone sum rules for the three-point
function of heavy baryons in order to determine the Isgur-Wise function.

\subsection*{Acknowledgments}
I want to thank my collaborators J.G.~K\"orner, A.A.~Pivovarov and
O.I.~Yakovlev for the continuing and fruitful collaboration. I would like to
thank the organizers of this conference for their hospitality. This work is
supported by a grant given by the DFG.

\end{document}